\documentclass{iopjournal}
\usepackage{amsmath,bm}
\usepackage{graphicx}
\begin{document}
\articletype{Note} 
\def\figscal{0.6}
\global\long\def\d{\mbox{d}}%
\global\long\def\va{\bm{a}}%

\title{Work needed to lift a mass in Schwarzschild spacetime}
\author{Anamta Asif$^1$\orcid{0009-0002-6823-1449} and Andrzej Czarnecki$^{2,*}$\orcid{0000-0003-1058-2529}}

\affil{$^1$School of Science and Engineering, Lahore University of Management
Sciences (LUMS), Opposite Sector U, D.H.A, Lahore 54792, Pakistan}
\email{25100098@lums.edu.pk}

\affil{$^2$Department of Physics, University of Alberta, Edmonton, Alberta,
Canada}
\email{andrzejc@ualberta.ca}

\affil{$^*$Author to whom any correspondence should be addressed.}


\begin{abstract}
  We determine the work needed to lift a mass from $r_2$ to 
  $r_1$ in Schwarzschild spacetime in three ways. First we use a winch
  and a light rope. Then we use a photon collision. Finally we let the
  mass decay into photons and reconstruct it at the higher
  radius. All three methods give the same local work at $r_1$,
  $W_1=mc^2[1-N(r_2)/N(r_1)]$, where $N(r)=\sqrt{1-r_s/r}$. Thus,
  independently of  $r_1$, the work tends to $mc^2$ as $r_2$ approaches the
  horizon.  The rope example also gives a useful general-relativistic
  version of a simple machine: 
a rope in equilibrium need not have
  equal local forces at its two ends. The redshift factor plays the
  role of the lever arm. 
For a real rope the energy conditions prevent the tension
from decreasing with height, so this lever advantage is confined to the idealization.
The examples are intended as teaching
  material for undergraduate relativity courses.
\end{abstract}

\section{Introduction}

The goal of this paper is to determine the work needed to lift a mass along the radial direction in Schwarzschild spacetime.  This geometry describes the vacuum gravitational field outside a spherical, nonrotating, uncharged mass. We are especially interested in the region just outside a black hole's event horizon. The lifted body and apparatus are assumed not to change the geometry (a ``test body'' approximation).

Our three examples are intended for upper-level undergraduates taking a first course in general relativity (GR). We assume familiarity with four-velocities, relativistic energy and momentum, and the use of a metric to compute proper times and distances, and scalar products.  The review in Section~\ref{sec:local-facts} also assumes some familiarity with the Schwarzschild metric, the geodesic equation, and covariant derivatives expressed through Christoffel symbols. Readers may take the results of that review as given when following the lifting examples.

This exercise has
several pedagogical virtues.

First, undergraduate students are used to mechanics problems involving
force and work. They also know simple machines, such as ropes, pulleys,
and levers. Such examples are common in introductory mechanics, but
less so in GR courses. We offer students
a worked GR example of the same kind: a winch with a rope lifts a mass.
The problem shows what is new in GR. Clocks at different radii run at
different rates, and a radial coordinate interval is not the same as the
proper length of a rope segment. The problem also shows what is familiar.
One can still compute work, but one must say whose work is being measured.

Second, we confirm the result with two other lifting schemes. One scheme
uses a collision with a photon. The other lets a massive particle decay
into photons and then recreates the particle at a different radial
coordinate. These examples give students several ways to solve the same
problem. They also separate the answer from the details of any one
mechanism.

Third, the winch and rope form a curious GR simple machine. In ordinary
mechanics, a massless rope in static equilibrium has the same tension at both ends.
In Schwarzschild spacetime this need not be true for local forces. Because
clocks run at different rates, the rope can be in equilibrium even when
the two local forces are different. This is similar in spirit to a lever.
For a lever, the ratio of forces is set by the ratio of arm lengths. For
the GR rope, the ratio of forces is set by the ratio of redshift factors.

The main result is simple. Let the lifting apparatus be held at
radius $r_1$. Let it lift a body of rest mass $m$ from $r_2$ to $r_1$,
where $r_s<r_2<r_1$, with $r_s$ denoting the Schwarzschild radius. Define
\begin{equation}
N_i=N(r_i)=\sqrt{1-\frac{r_s}{r_i}},
\qquad
r_s=\frac{2GM}{c^2}.
\end{equation}
The local work supplied by the apparatus at $r_1$ will be shown to be
\begin{equation}
W_1=mc^2\left(1-\frac{N_2}{N_1}\right).
\label{eq:W1main}
\end{equation}
By local work at radius $r$ we mean work measured by a static observer
whose apparatus is held at that same radius.  The observer uses their
own proper time, local rods, and local energy units.  If a local force
$ F$ acts through a local proper displacement $\d\ell$, then
\begin{equation}
  \d W_{\rm loc}(r)= F\,\d\ell .
\end{equation}
$W_{\rm loc}$ can be determined  for example by the number of nuclei that
undergo fission at $r$.

A consequence of Eq.~\eqref{eq:W1main} is 
\begin{equation}
W_1\to mc^2
\quad\text{as}\quad r_2\to r_s^+.
\label{eq:r2=rs}
\end{equation}
A counterintuitive meaning of Eq.~\eqref{eq:r2=rs} is worth
spelling out.  No matter to what $r_1$ the mass is lifted, the local
work at $r_1$ is always  $mc^2$ if $r_2 \to r_s$.

Thus a winch held near the horizon and a winch held far away both supply
the same local energy when the body is retrieved from the horizon limit
to rest beside the winch.  This does not mean that the same work is
obtained after comparison at a common radius.  Local work needed at
$r_1$ is redshifted by the factor $N_1/N_0$ when it is supplied by 
a static observer at $r_0$ (this will be especially clear in the
photon-collision scheme described below).  In those common units the
near-horizon limit is $mc^2N_1/N_0$, which depends on the final radius
$r_1$.  The independence of $r_1$ occurs only when each observer
measures the work locally at the radius where the apparatus is held.

This conclusion is part of standard energy bookkeeping near black holes.
It is implicit in classic lowering-box and rope arguments. Bekenstein used
lowering-box thought experiments in his discussion of the generalized
second law \cite{Bekenstein:1974ax}. Gibbons studied the work obtained by
lowering a particle on a rope into a black hole or its ergosphere
\cite{Gibbons:1972qjv}. Unruh and Wald studied lowered boxes, acceleration
radiation, and black-hole mining
\cite{Unruh:1982ic,UnruhWald1983Mining}.

Later work studied strings as channels through which black holes can lose
energy \cite{Lawrence:1993sg,Frolov:2000kx}, with thermally excited
transverse modes carrying energy outward along the string. 
In contrast, the rope in the present paper is a classical
device used to raise a mass.

Force language has been used in pedagogical discussions of
Schwarzschild spacetime.  Ridgely derives force formulae in general
relativity from local energy-momentum conservation.  He obtains the local weight of an observer held at fixed radius
and recovers the Newtonian result in the weak-field limit
\cite{Ridgely2010}.  Doughty studies the proper acceleration of static
observers, meaning the acceleration read by an accelerometer carried by
the observer, and uses this quantity to physically interpret
surface gravity: the finite horizon limit of the local acceleration after
multiplication by the redshift factor \cite{Doughty1981}.  We shall
use those results concerning the local weight and gravity in Section \ref{sec:string}.  Jonsson gives
an intuitive discussion of real thrust forces near a black hole.  His
main Schwarzschild example is the centrifugal-force paradox: inside the
photon orbit radius, increasing the speed of a circularly moving rocket
increases the outward thrust needed to maintain the circular orbit
\cite{Jonsson2006}.

 Brown discusses
the material-strength limits of idealized black-hole mining devices
\cite{Brown:2012un}.  
MacLaurin
studies extended cables in static, spherically symmetric spacetimes.  For
a stationary cable, the tension changes with height by a redshift
factor \cite{MacLaurin:2019lew}.  This redshift is a key element in our study.
The present paper is meant as a complement to these works.  We fix the apparatus at a
finite radius $r_1$ and compute the local work $W_1$ done there.  The
rope calculation lets students use the familiar idea of force transmitted
through a string.  It also shows the new relativistic ingredient: the
redshift factor $N=d\tau/dt$ plays the role of an effective lever arm.
A local force near $r_2$ can become arbitrarily large,
while the work done by the apparatus at $r_1$ has the finite limit
$mc^2$ as $r_2\to r_s^+$.

A final pedagogical virtue of the exercise is that it clarifies a related
near-horizon confusion. A static observer near the horizon sees a
freely falling body passing it
at almost the speed of light. One may therefore be tempted to say that a
body near the horizon sits at the bottom of an infinite potential well.
For example, one pedagogical account says that ``the energy required
becomes infinite if the particle is placed exactly on the event horizon''
\cite{Rezzolla:2023zcj}. This is a useful warning:  a static observer
cannot be placed on the horizon. However, such wording can 
obscure the fact that, if a winch held at a finite radius $r_1$ lifts a body
from $r_2>r_s$ to $r_1$, the local work done by the winch has a finite
limit as $r_2\to r_s^+$. The calculations below show this finiteness.

The paper is organized as follows. Section~\ref{sec:local-facts} reviews
two useful results in Schwarzschild spacetime.
Sections~\ref{sec:string}, \ref{sec:photon},
and \ref{sec:pion} derive Eq.~\eqref{eq:W1main}  using the three lifting
protocols.  Section \ref{sec:Killing} uses the conservation of energy measured at infinity to explain why the three scenarios must give the same result.

Throughout this paper, $r$ and $t$ denote standard Schwarzschild
coordinates: the areal radius coordinate $r$ is such that a sphere at constant $r$ has the area
$4\pi r^2$ and $t$ is the time measured by clocks at $r\to \infty$. 

\section{Useful results in Schwarzschild spacetime: a review}
\label{sec:local-facts}
In this Section we review basic calculations in  Schwarzschild
spacetime. For more details see any introductory GR textbook, for
example \cite{MooreGR2015,schutz2022first}.

\subsection{Freely falling mass}

Consider mass $m$ freely falling from initial rest at $r=r_{1}$.
We parameterize its world line by its proper time $\tau$,
\begin{equation}
c^{2}\d\tau^{2}=-g_{tt}\d t^{2}-g_{rr}\d r^{2}=N^{2}c^{2}\d t^{2}-\frac{\d r^{2}}{N^{2}},
\end{equation}
where $N\left(r\right)=\sqrt{1-r_{s}/r}$, so that $g_{tt}=-N^{2}c^{2}$
and $g_{rr}=N^{-2}$. We use $u^{\mu}=\d x^{\mu}/\d\tau$
for the 4-velocity. Its square is
\begin{equation}
-c^{2}=u^{2}=g_{tt}\left(u^{t}\right)^{2}+g_{rr}\left(u^{r}\right)^{2}.\label{eq:unit4vel}
\end{equation}
The $t$-component of the geodesic equation is
\begin{equation}
\frac{\d}{\d\tau}\left(g_{tt}u^{t}\right)=0.
\end{equation} 
We conclude that the constant $g_{tt}u^{t}$ equals $-c^{2}N_{1}$
and
\begin{equation}
u^{t}(r) =\frac{N_{1}}{N(r)^{2}},\qquad u^{r}(r)=-c\sqrt{N_{1}^{2}-N(r)^{2}}.
\label{eq:utur}
\end{equation}
Consider now a static observer $O$ at $r=r_{2}$, with $r_{s}<r_{2}<r_{1}$.
(We will be especially interested in $r_{2}$ very close to the Schwarzschild
radius $r_{s}$.) The observer uses an orthonormal frame; denote the
unit vector in the radial direction by $o_{r}$. In $O$'s rest frame,
the falling mass has 4-velocity $\left(\gamma c,-\gamma v\right)$ where
$\gamma=1/\sqrt{1-v^{2}/c^{2}}$ (since we consider a radial motion
at constant spherical angles, we display only the time and radial
components of 4-vectors). To determine the value of the speed $v$
seen by $O$, consider the invariant product
\begin{equation}
-\gamma v=o_{r}\cdot u=g_{rr}\left(o_{r}\right)^{r}u^{r}.\label{eq:or}
\end{equation}
Since $o_{r}$ is a unit vector, we have
\begin{equation}
1=g_{rr}\left[\left(o_{r}\right)^{r}\right]^{2},
\end{equation}
so $\left(o_{r}\right)^{r}=1/\sqrt{g_{rr}}$ and 
\begin{equation}
-\gamma v=\sqrt{g_{rr}}u^{r}=-c\sqrt{\frac{N_{1}^{2}-N_{2}^{2}}{N_{2}^{2}}}.
\end{equation}
We solve this for $v$ finding $v=c\sqrt{1-N_{2}^{2}/N_{1}^{2}}$.
We see that $v$ tends to the speed of light as $N_{2}$ becomes small
when $r_{2}$ nears $r_{s}$. 

This is the origin of the occasional confusion we mention in this
paper. The local observer, seeing the mass approaching the speed of
light and therefore its kinetic energy growing to infinity, may
wrongly conclude that the event horizon resembles the bottom of an
infinite potential well, and that the work needed to retrieve a mass
resting at $r_{2}$ tends to infinity as $r_{2}\to r_{s}$.

We shall demonstrate that redshift factors ensure the finiteness of
the work done  at $r_1$. The divergent
kinetic energy measured by hovering observers near the horizon is a
warning that local quantities measured at different radii must not be
confused. The divergence does not imply that an infinite work is needed
to retrieve a mass.

\subsection{Force needed to hover}

We now determine the force needed to keep the observer $O$ at constant
radial coordinate $r$. For a body at rest, the only nonzero component
of the 4-velocity is $u^{t}=1/N$, see Eq.~(\ref{eq:unit4vel}). The 4-acceleration, given by $a^{\mu}=u^{\nu}\nabla_{\nu}u^{\mu}$,
is
\begin{equation}
a^{\mu}=u^{t}\left(\partial_{t}u^{\mu}+\Gamma_{tt}^{\mu}u^{t}\right)=\left(u^{t}\right)^{2}\Gamma_{tt}^{\mu},
\end{equation}
where $\Gamma_{\alpha\beta}^{\mu}$ denote Christoffel symbols. The
only nonzero contribution comes from
\begin{equation}
\Gamma_{tt}^{r}=-\frac{g^{rr}}{2}\partial_{r}g_{tt}=\frac{c^{2}N^{2}}{2}\partial_{r}N^{2}=\frac{c^{2}N^{2}}{2}\frac{r_{s}}{r^{2}},
\end{equation}
leading to
\begin{equation}
a^{r}=\frac{c^{2}r_{s}}{2r^{2}}.
\end{equation}
The proper acceleration (the accelerometer reading \cite{Doughty1981}) of a static observer
is found analogously to Eq.~(\ref{eq:or}),
\begin{equation}
g(r)=o_{r}\cdot
a=\frac{1}{N}a^{r}=\frac{c^{2}r_{s}}{2r^{2}\sqrt{1-\frac{r_{s}}{r}}}.
\label{eq:g}
\end{equation}
This locally perceived acceleration diverges when $r\to r_{s}$,
so that the force needed to keep a mass stationary near the horizon
diverges as well. Despite that divergent force, we shall see that
the work done by an apparatus held at $r=r_1$ has a finite limit as
$r_2\to r_s^+$. 

\section{Lifting with a light rope}\label{sec:string}

\begin{figure}[htb]
\centering
\includegraphics[scale=\figscal]{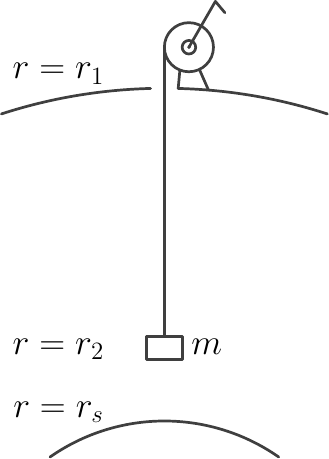}
\caption{Mass $m$ is lifted from $r=r_2>r_s$ to $r=r_1$ by a light,
negligibly extensible rope. The quantity computed in Section \ref{sec:string} is
the local work done by the winch at $r=r_1$.\label{fig:winch}}
\end{figure}

 We consider a mass $m$ initially held at
$r=r_2>r_s$ and 
lifted to $r=r_1$; see Fig.~\ref{fig:winch}. %
The body is attached to an idealized massless, negligibly extensible
rope. Since the winch is static at $r_1$,
the quantity naturally computed here is the local work $W_1$ supplied at
that radius.  

For an ideal massless rope in quasistatic equilibrium, the impulse
transmitted through the rope during a coordinate-time interval $\d t$
is the same at every radius. The local proper times satisfy
$\d\tau(r)=N(r)\,\d t$ and $\d\tau_1=N_1\,\d t$, so the corresponding
local impulses are $F(r)\,\d\tau(r)$ and $F_1\,\d\tau_1$. Hence
\begin{equation}
F_1N_1=F(r)N(r).
\label{eq:lever}
\end{equation}
Equation~\eqref{eq:lever} may be viewed as a gravitational analogue of a
lever law. The redshift factor $N=d\tau/dt$ plays the role of an
effective arm length. Static worldlines deeper in the field have smaller
$N$, so a larger local force there can be balanced by a smaller local
force at larger radius.  
 
%Our justification of Eq.~\eqref{eq:lever} is meant to appeal to
%intuition.  It can be made rigorous by invoking the stress-energy
%conservation (see \cite{MacLaurin:2019lew} and \cite[end of
%Sect.~2.1]{Czarnecki:2026bjb}, and Sect.~{sec:Killing} below).
Our justification of Eq.~\eqref{eq:lever} is meant to appeal to
intuition. It can be made rigorous: conservation of the stress-energy
tensor of a static massless rope implies that the redshifted tension
$F(r)N(r)$, the force conjugate to the timelike Killing vector
$\partial_t$ (that is, the force expressed in the units of energy at
infinity used in Section~\ref{sec:Killing}), is constant along the
rope (see \cite{MacLaurin:2019lew} and \cite[end of
Sect.~2.1]{Czarnecki:2026bjb}).

The lower-end tension $F(r_2)=mg(r_2)$ diverges as $N_2^{-1}$ when
$r_2\to r_s^+$. Within the massless-rope model, the winch force remains
finite for fixed $r_1>r_s$:
\begin{equation}
F_1=\frac{mc^2r_s}{2r_2^2N_1}
\longrightarrow \frac{mc^2}{2r_sN_1}.
\label{eq:winchforce-limit}
\end{equation}
%The horizon limit thus requires a rope that can sustain arbitrarily
%large local tension. A real rope has finite strength and its own
%weight, which must also be supported \cite{Brown:2012un}. The finite
%winch force above relies on neglecting that weight.
The horizon limit thus requires a rope that can sustain arbitrarily
large local tension. In fact, the lever advantage itself is an
artifact of the massless idealization. For a static rope with proper
mass per unit length $\mu$ and tension $F$, stress-energy conservation
gives \cite[Eq.~(5)]{Brown:2012un}
\begin{equation}
\frac{\d F}{\d\ell}=\left(\mu c^{2}-F\right)\frac{g(r)}{c^{2}},
\label{eq:massiverope}
\end{equation}
where $\ell$ is the proper radial distance, $\d\ell=\d r/N(r)$; for
$\mu=0$ this reduces to Eq.~\eqref{eq:lever}. The energy conditions
require $F\le\mu c^{2}$ for any real rope \cite{Brown:2012un}, so its
tension cannot decrease with height: $F_1\ge F(r_2)=mg(r_2)$, and the
winch force diverges together with the local weight as
$r_2\to r_s^+$. The lever law~\eqref{eq:lever} and the finite
limit~\eqref{eq:winchforce-limit} therefore hold only for the
idealized massless rope. The local work $W_1$ derived below is not
affected: by the energy conservation of Section~\ref{sec:Killing}, a
massive rope only adds the work needed to retrieve the rope itself.

When the body moves from $r$ to $r+\delta r$, the proper length
$\delta\ell$ of the rope removed
by the winch equals the proper radial interval,
\begin{equation}
\delta \ell=\frac{\delta r}{N(r)},
\end{equation}
so the local work supplied at $r_1$ is
\begin{equation}
\delta W_1=F_1\,\delta \ell=\frac{F(r)}{N_1}\,\delta r.
\end{equation}
The force needed to keep mass $m$ static at $r$ is $F(r) = mg(r)$ \cite{Ridgely2010},
where the proper acceleration $g(r)$ was found in Eq.~\eqref{eq:g},
\begin{equation}
F(r)=mg(r),\qquad g(r)=\frac{c^{2}r_s}{2r^{2}N(r)}.
\end{equation}
We thus arrive at our main result, advertised in Eq.~\eqref{eq:W1main},
\begin{equation}
W_1=\frac{mc^{2}r_s}{2N_1}\int_{r_2}^{r_1}\frac{\d r}{r^{2}N(r)}
=mc^{2}\left(1-\frac{N_2}{N_1}\right),
\label{wrope_local}
\end{equation}
because $\d N/\d r=r_s/\left(2r^{2}N(r)\right)$.
Therefore, for fixed
$r_1$,
\begin{equation}
W_1\to mc^{2}
\qquad \text{as} \qquad r_2\to r_s^+.
\end{equation}

\section{Lifting with photon scattering}\label{sec:photon}

\begin{figure}[htb]
\centering
\includegraphics[scale=\figscal]{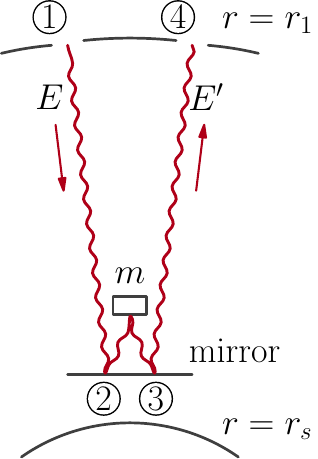}
\caption{(1) A photon of local energy $E$ is emitted at $r=r_1$; (2) it is reflected by
 a stationary mirror and scatters elastically from the mass $m$ at $r=r_2$.
  After (3) reflecting from the mirror again, it (4) returns to $r=r_1$ with local energy $E^\prime $. If the mass just reaches
 $r_1$ and comes to rest there, the work supplied at $r_1$ is $E-E^\prime $. The mirror can be imagined as  being supported by a rocket engine and  very massive so the photon does not lose energy to it.}
\label{fig:photon}
\end{figure}

We now recover the same result  from scattering kinematics. In
this section we set $c=1$.
A photon of local energy $E$ is emitted at $r=r_1$, reflected by a
heavy, stationary mirror below the body, scatters elastically from the
mass $m$ at $r=r_2$, is reflected again by the same mirror, and
returns to $r_1$ with local energy $E^\prime $; see
Fig.~\ref{fig:photon}. 

The local work supplied by the apparatus at $r_1$
is the same as the energy lost by the photon,
\begin{equation}
W_1^{(\gamma)}=E-E^\prime .
\end{equation}

Let $E_i$ and $E_f$ be the incoming and outgoing photon energies in the
instantaneous rest frame of the mass at $r_2$. Since photon energies
redshift in proportion to $1/N$,
\begin{equation}
E_i=E\frac{N_1}{N_2},
\label{ei}
\qquad
E^\prime =E_f\frac{N_2}{N_1}.
\end{equation}
To just reach $r_1$ and stop, the mass must leave the collision with the
outward momentum of a radial geodesic whose turning point is at $r_1$.
As we found in Eq.~\eqref{eq:utur},
\begin{equation}
\left(\frac{dr}{d\tau}\right)^2=N_1^2-N(r)^2,
\end{equation}
so at $r_2$ the required local momentum is
\begin{equation}
p_{\rm esc}=m\,\frac{1}{N_2}\left.\frac{dr}{d\tau}\right|_{r_2}
=\frac{m}{N_2}\sqrt{N_1^2-N_2^2}.
\label{eq:pescphoton}
\end{equation}
In the same local frame, one-dimensional elastic scattering gives
\begin{equation}
m+E_i=\sqrt{m^2+p_{\rm esc}^2}+E_f,
\qquad
p_{\rm esc}=E_i+E_f.
\end{equation}
Subtracting these relations,
\begin{equation}
E_i-E_f=\sqrt{m^2+p_{\rm esc}^2}-m.
\end{equation}
Using Eq.~\eqref{eq:pescphoton},
\begin{equation}
\sqrt{m^2+p_{\rm esc}^2}=m\frac{N_1}{N_2},
\end{equation}
and therefore the work at $r_1$ determined as the energy lost by the
photon is
\begin{equation}
W_1^{(\gamma)}=E-E^\prime =\frac{N_2}{N_1}(E_i-E_f)
=m\left(1-\frac{N_2}{N_1}\right).
\label{eq:Wphoton}
\end{equation}
This is Eq.~\eqref{eq:W1main} with $c=1$.

\section{Decay into photons and reconstruction}\label{sec:pion}

A third realization replaces the particle by radiation near $r_2$ and
reconstructs it at $r_1$ (this is possible for particles that decay
into photons, for example a neutral pion, the Higgs boson, or a
positronium atom). If a particle of rest mass $m$ converts at
$r_2$ into photons that all escape, the radiation arriving at $r_1$ carries local energy
$(N_2/N_1)mc^2$ (see Eq.~\eqref{ei}). Recreating the particle at rest at $r_1$ therefore
requires additional local energy
\begin{equation}
W_1^{\text{rad}}=mc^2\left(1-\frac{N_2}{N_1}\right),
\end{equation}
in agreement with Eq.~\eqref{eq:W1main}.

\section{Killing energy point of view}
\label{sec:Killing}
In this Section, outside our main story line, we present a more general (but more abstract) point of view.  It is not accidental that the three scenarios we considered gave the same results for work. There is an underlying principle: conservation of the so-called Killing energy (or ``energy at infinity''; see \cite[p.~270]{schutz2022first}). In this Section we again use $c=1$.

If the metric is independent of a coordinate $x^a$, then the corresponding covariant momentum component $p_a$ is conserved along any geodesic. The Schwarzschild metric is independent of $t$, so $p_t$ is conserved. We define its rescaled version,
\begin{equation}
  \label{eq:KilEn}
  e\equiv -{p_t \over m} = -g_{tt} {\d t \over \d \tau}=\left( 1-{2GM \over r}\right) {\d t \over \d \tau},
\end{equation}
where $m$ is a particle's rest mass and $\tau$ its proper time; $e\cdot m$ is called the Killing energy of the particle. 

A local energy $E_{\text{loc}}$ at radius $r$ corresponds to the energy measured at $r\to \infty$ equal $E_\infty = N(r) E_{\text {loc}}$. For example, a body of mass $m$ at rest at $r$ has  $E_\infty(r) = N(r)m$. With no losses or other net energy changes, conservation gives\
\begin{equation}
  \label{eq:cons}
  N_1 W_1 = E_\infty(r_1 ) -  E_\infty(r_2 ) = m(N_1 - N_2). 
\end{equation}
For this reason, all three scenarios must give the same local work.

The purpose of our paper has been to supplement this somewhat abstract argumentation with concrete protocols.

\section{Conclusions}
\label{sec:conclusion}

The three retrieval scenarios studied here are highly idealized. We work
in Schwarzschild spacetime in the test-body limit. We also neglect the
material limitations of real ropes, winches, and mirrors; those limitations
are important in black-hole mining thought experiments \cite{Brown:2012un}.

For the mass-retrieval problem, the most relevant quantity is the local work done
by the apparatus held at $r_1$. For a quasistatic transfer from $r_2$ to
$r_1$, with $N_i=N(r_i)$, this work is
\begin{equation}
W_1 = mc^2\left(1-\frac{N_2}{N_1}\right).
\end{equation}
The three realizations studied here --- a light rope, photon scattering,
and decay followed by reconstruction --- all lead to the same result.

As shown in Section~\ref{sec:Killing}, this agreement follows from the
conservation of the Killing energy (energy at infinity),
$N_1W_1=mc^2(N_1-N_2)$, which fixes the local work at $r_1$
independently of the lifting mechanism.

The rope picture offers the clearest mechanical intuition: for an idealized massless rope, Eq.~\eqref{eq:lever} acts as a gravitational analogue of a lever law, with
the redshift factor $N=d\tau/dt$ playing the role of an effective arm
length. Near the horizon the local force needed to hold the body diverges.
But during the lift the redshift factor and the proper radial distance also
matter. Their product gives a finite local work.

The lever advantage, unlike the work, holds only for the massless-rope
idealization: a real rope obeys the energy-condition bound
$F\le\mu c^2$ \cite{Brown:2012un}, so its tension cannot decrease
with height and the winch force diverges together with the local
weight (Section~\ref{sec:string}).

Thus, for fixed $r_1$, retrieving a body from $r_2>r_s$ to $r_1$ has the
finite near-horizon limit
\begin{equation}
W_1\to mc^2
\qquad \text{as} \qquad
r_2\to r_s^+.
\end{equation}
We
stress that the horizon itself is excluded, since no static worldline
exists there. We hope that the comparison of the three scenarios presented
in this paper will be useful as teaching examples for near-horizon energy
bookkeeping.

\subsection*{Acknowledgements}
This research was supported by the Natural Sciences and Engineering
Research Council of Canada (NSERC).

%\bibliography{/Users/czar/Dropbox/pro/Tables/Archive/ac}
%\bibliographystyle{/Users/czar/Dropbox/pro/Tables/Archive/Bibtex/andrzej}

\end{document}